\documentclass[twocolumn,showpacs,%
  nofootinbib,aps,superscriptaddress,%
eqsecnum,prd,notitlepage,showkeys,10pt,nobalancelastpage]{revtex4-1}

\usepackage{amssymb}
\usepackage{amsmath}
\usepackage{graphicx}
\usepackage{dcolumn}
\usepackage{hyperref}
\usepackage{comment}
\usepackage{longtable}
\usepackage{booktabs}
\usepackage{cleveref}
\usepackage{balance}



\begin{document}

\title{Fitness advantages, prediction of resistances and protocol design:\\exploiting the synergy between mathematical modelling and statistics}
\author{Giulia Chiari}
\affiliation{BCAM - Basque Center for Applied Mathematics\\Mazarredo Zumarkalea, 14, Abando, 48009 Bilbao, Bizkaia, Spain}

\maketitle

\section{Introduction}
Cancer progression is governed by evolutionary principles, where heterogeneous cell populations compete under selective pressures \cite{vanderheijden2019}. Clonal selection (the preferential growth of subpopulations with fitness advantages) drives tumour growth and treatment response. Within this process, two phenomena are central: clonal expansion, the rise of a dominant subclone, and clonal inversion, the replacement of a once-dominant clone by a previously minor population, often triggered by therapy or microenvironmental changes. Both reflect how cellular variability interacts with selection to shape tumour composition \cite{ding2012}.

A key challenge is to identify which tumour features most strongly influence clonal evolution. Proliferation rate, repair capacity, and epigenetic plasticity can accelerate adaptation and underlie treatment resistance. Therapy itself acts as an evolutionary force, altering selection and fitness advantage.  Radiotherapy, one of the most widely used strategies in cancer treatment, can be personalized by designing protocols that vary in total dose, fractionation, and timing. The goal is not only to maximize cancer cell killing, but also to influence evolutionary dynamics to limit the clonal expansion of resistant subpopulations \cite{supiot2010}.

In this work, we present a novel combination of mathematical modelling and statistical inference analysis that proves to be an essential tool in the integration of experimental observations to quantify selection, identify hidden drivers, and predict resistance trajectories \cite{colson2025}. This strategy can ultimately guide the design of treatment protocols to anticipate and counteract evolutionary escape.

\section{Methods}
We consider a cancer mass composed of $N$ clones, evolving according to resource availability and treated with radiotherapy \cite{dimitriou2024}. 
We denote with $x_i$ the number of cancer cells in the $i$-th population, where $i$ is determined by a combination of proliferation rate $\gamma_i$, radiosensitivity $\alpha_i$ in the sense of the LQ formulation for radiation damage \cite{mcmahon2018}, and $\frac{\alpha_i}{\beta_i}$ ratio establishing the sensitivity to fractionation \cite{vanleeuwen2018}.

We model these dynamics as follows:
$$\dfrac{d x_i}{dt} = (1-\mu) P_i +\mu S_i \sum_{k=0}^NP_k   
$$
where
$$P_i  = \gamma_i \left(1-\dfrac{\sum_{j=0}^N \gamma_j x_j}{K}\right) x_i$$
is the proliferative term and $S_i$ is the $i$-th component of an N-dimensional time-dependent vector generated with Dirichlet distribution (epigenetic redistribution \cite{zheng2017}). Here $K$ is the resource availability (i.e. glucose) and we consider its consumption by each population to be proportional to its proliferative activity. Moreover, $\mu$ represents the epigenetic instability, considered as environment-dependent and acting equally on all populations.

Radiotherapy is modelled as a set of events, where single dose administration $d$ leads to a clone-specific survival fraction $$SF_i = e^{-\alpha_i d -\beta_id^2}.$$
Specifically, we consider the timing and dosage of individual fractions, determined by combinations of total dose (60, 72, or 84 Gy), timing (A–accelerated, S–standard, P–protracted), and fractionation scheme (O–hypofractionated, R–regular, E–hyperfractionated).

We analyse the evolutive dynamics of the system and its equilibria, concentrating on the coexistence steady state, represented by line $K-\sum_{j=0}^N \gamma_j x_j=0$, where the percentage of every $x_i$ on the total depends on $\gamma_i$, $\mu$ and the initial conditions. Radiotherapy events reduce the population and alter the contribution of different clones.

We simulate the model adopting the uniformity rate of the initial data distribution among the clones, the epigenetic instability, and the choice of protocol as random variables. We generate an in silico database of 255,150 entries, with variables defined and described in TABLE \ref{tab:1}, \ref{tab:2}).

We perform statistical analyses to investigate the following questions:
\begin{itemize}
    \item[Q1] How do evolution and radiotherapy influence clonal selection?
    \item[Q2] Are there any biological characteristics that influence the performance of the protocols?
    \item[Q3] Which protocol is optimal? Does it change according to specific patient needs?
    \item[Q4] Is this framework consistent with experimental evidence?  
\end{itemize}

The main techniques used include Ordinary Least Squares (OLS) models with interaction terms to evaluate the joint and individual effects of treatment and biological related variables, analysis of variance, rank correlation for pairwise relationships, and quantitative comparisons of stratified means to explore systematic effects of specific factors. Finally, unsupervised clustering using the k-means algorithm on selected variables is applied to identify recurring patterns in the multidimensional dataset.

\section{Results}
We present results addressing the questions introduced above, with references supporting the corresponding biological evidence.

\begin{figure*}
    \centering
    \includegraphics[width=0.8\linewidth]{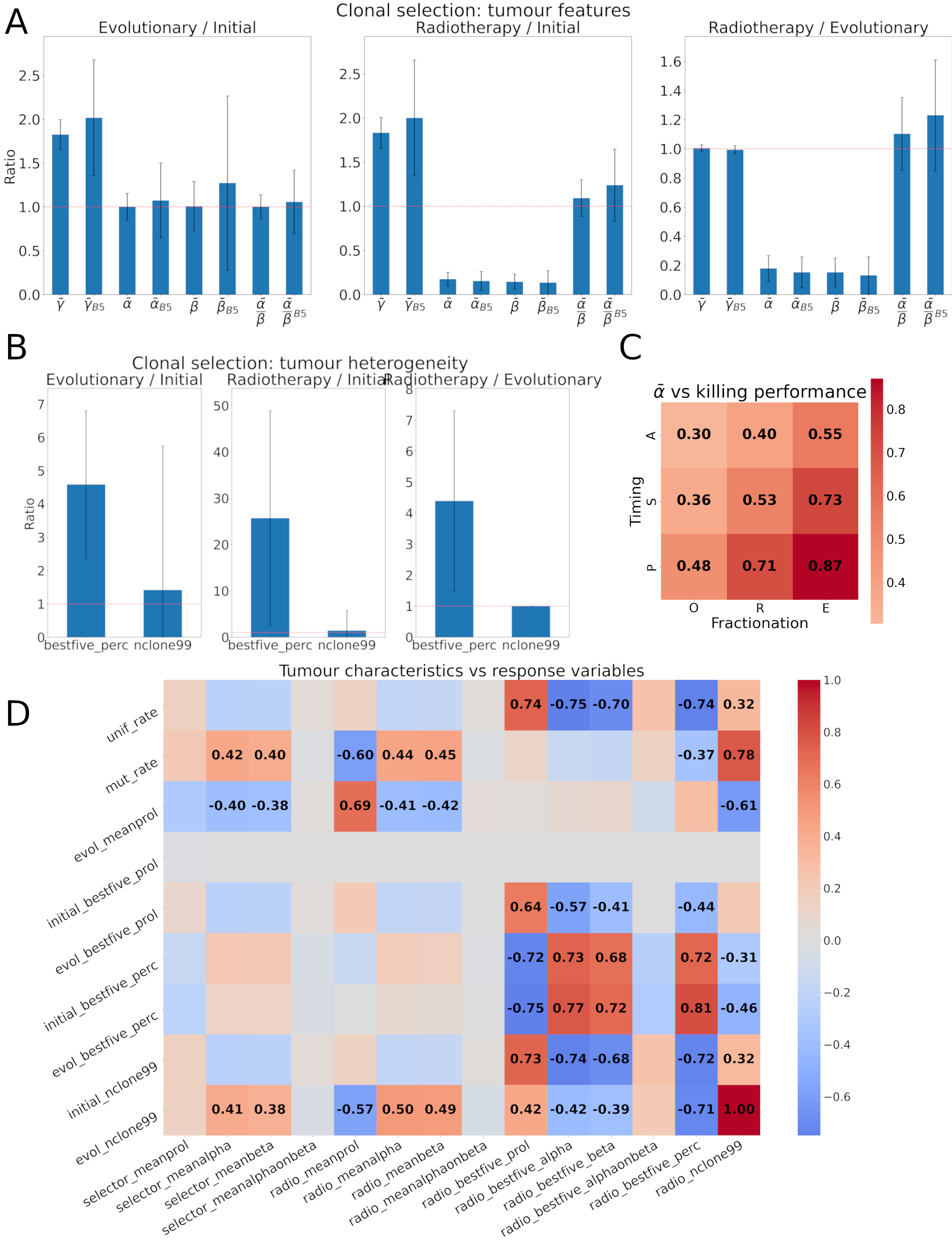}
    \caption{Analysis of clonal expansion and selection. (A) Analysis of the means of radio-sensitivity parameters $\alpha$,$\beta$,$\dfrac{\alpha}{\beta}$ and of the proliferation parameter $\gamma$, evaluated on the total cancer mass and on the top five clones by abundance, considering three different times: starting point, detection (cancer mass selected by evolutive dynamics), and recurrence (cancer mass selected by radiotherapy) for first-line therapy. Here we represent the ratio between the means considering all the combinations among the times considered. B) Analysis of mean percentage on the total mass represented by top five clones by abundance and of the minimum number of clones to reach the $99\%$ of the cancer mass. We mantain the same analysis structure as in (A). (C) Analysis on the influence of $\bar{\alpha}$ (mean value) on cancer cell killed percentage on the total mass. (D) Analysis of the correlation between tumour features and protocol outcomes in term of clonal selection.}
    \label{fig:1}
\end{figure*}

With respect to the initial distribution of epigenetic traits within the population, the selective dynamics induced by evolutionary stress favours highly proliferative clones, whereas therapy exerts selective pressure toward radioresistant clones \cite{mcabee2019} (FIG. \ref{fig:1}A). With the current modelling, therapy does not interfere with the evolutionary selection acting on proliferative traits.
Another important selection, exerted both by inter-clonal competition and, more prominently, by therapy, concerns population heterogeneity, which is markedly reduced, leading the tumour mass to be represented by fewer clones \cite{morrissy2016} (FIG. \ref{fig:1}B).

Regarding the impact of tumour characteristics on therapeutic outcome, the most determining factor is $\bar{\alpha}$, whose influence becomes more significant in prolonged and hyperfractionated treatments, progressively diminishing as we move toward accelerated and hypofractionated regimens (FIG. \ref{fig:1}C). This trend is consistent with clinical practice, which favours the latter for tumours characterized by a low $\dfrac{\alpha}{\beta}$ ratio \cite{brand2022}.

Interestingly, the specific features of the initial tumour mass do not affect its evolution \cite{morrissy2016}, whereas population heterogeneity is highly relevant: highly diverse tumours correlate with greater post-relapse selection of highly proliferative and poorly radiosensitive traits, while maintaining higher overall heterogeneity (FIG. \ref{fig:1}D). On the other hand, high epigenetic instability can be beneficial, preventing favoured clones from expanding. Tumours exhibiting stronger selection for proliferative rate during evolution drive a faster emergence of radio-resistance under radiotherapy \cite{walens2020,lamprecht2017}.

\begin{figure*}
    \centering
    \includegraphics[width=0.8\linewidth]{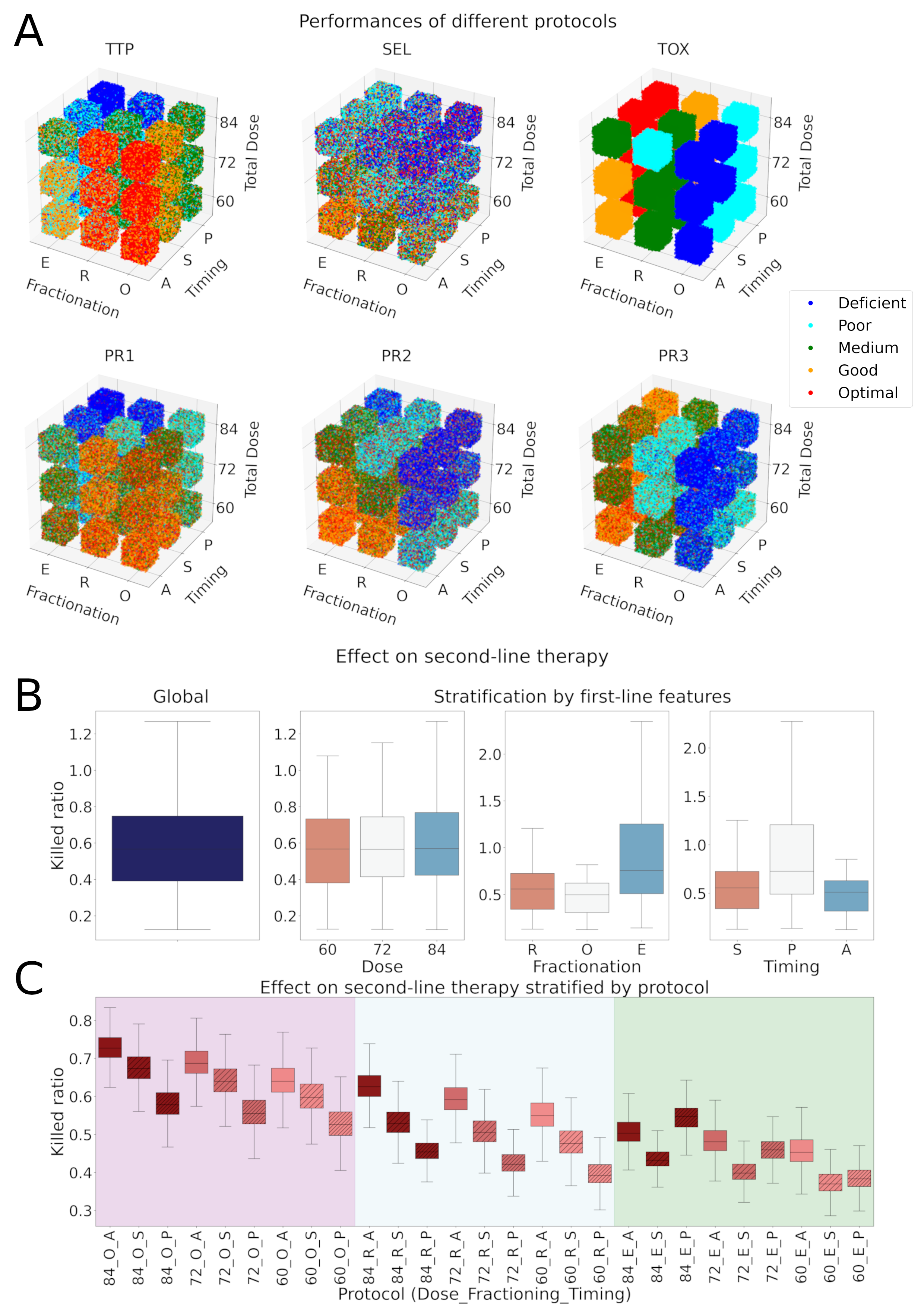}
    \caption{Analysis of protocols. (A) All experiments are grouped by first-line protocol features. In first row, protocols are evaluated by TTP (time to progression), SEL (ratio between mean values for $\alpha$ considering post- and pre-treatment state) and TOX (toxicity, evaluated using BED theory, adjusted to account for damage repair during the interval between doses). In second row, use weighted combinations of the three indices described above to define three representative patient profiles: PR1 corresponds to a robust patient aiming for eradication while tolerating high doses; PR2 represents a patient in good condition but susceptible to side effects at high doses, aiming to prolong relapse-free intervals while preserving a sensitive cell population for potential subsequent treatment; finally, PR3 represents a high-risk patient who can only tolerate moderate-intensity therapies, focusing the treatment strategy on repeated cycles while seeking to maintain efficacy across multiple sessions. (B) Evaluation of cancer cell killing performance of the same protocols adopted as second- vs first-line treatment (we plot the ratio of the mean values second/first line), globally considered and stratified by first line features. (C) Analysis of the decrease in cancer cell killing potential of protocols used as second-line treatments, stratified by dose, fractionation and timing combinations (background colours characterise fractionation, boxplot colours define total dose, patterns identify timing).}
    \label{fig:2}
\end{figure*}

Considering dose painting optimization, our model indicates that combinations characterised by high doses, hypofractionation, and accelerated administration maximize time to progression, albeit at the cost of increased toxicity \cite{sinzabakira2024}, which is instead minimized by low-dose, hyperfractionated, and prolonged protocols (FIG. \ref{fig:2}A). Hyperfractionation and low doses also reduce the selection pressure toward radio-resistance \cite{haslett2014}.

Subsequently, we present a set of optimal protocols with the corresponding variance for three representative patient profiles (described in the caption) \cite{karaca2014}.

We further observe that the same protocol (FIG. \ref{fig:2}B), when used as a second-line treatment, leads to a marked deterioration in performance \cite{xu2024}. The most pronounced declines occur for treatments that follow a first-line hypofractionated and accelerated regimen. Prolonged, hyperfractionated, and low-dose protocols appear to be more sensitive to performance degradation \cite{takeda2021}(FIG. \ref{fig:2}C).

\section{Discussion}
In this work, we present a novel framework combining mathematical modelling and statistical inference to study radiotherapy optimization strategies. We characterise clonal selection and expansion driven by evolutionary dynamics and therapy, highlighting biological factors influencing treatment outcomes. Furthermore, we identify optimal therapeutic strategies tailored to prototype patients consistent with clinical practice. All findings align with experimental data and clinical evidence, demonstrating the framework’s potential utility to support clinical therapy planning.

\bibliographystyle{unsrt}
\bibliography{references}


\begin{table*}[]
\begin{tabular}{l|l|l|l|l|l|l|l|l|l|l|}
\cline{2-11}
                                                  & count  & mean    & std     & min     & 25\%    & 50\%    & 75\%    & max    & median  & range  \\ \hline
\multicolumn{1}{|l|}{unif\_rate}                  & 255150 & 22.22   & 39.07   & 0.01    & 0.1     & 1       & 10      & 100    & 1       & 99.99  \\ \hline
\multicolumn{1}{|l|}{mut\_rate}                   & 255150 & 0.02    & 0.03    & 0.00001 & 0.00005 & 0.001   & 0.02    & 0.1    & 0.001   & 0.1    \\ \hline
\multicolumn{1}{|l|}{initial\_meanprol}           & 255150 & 0.03    & 0.00215 & 0.00899 & 0.03    & 0.03    & 0.03    & 0.04   & 0.03    & 0.03   \\ \hline
\multicolumn{1}{|l|}{initial\_meanalpha}          & 255150 & 0.1     & 0.00868 & 0.03    & 0.1     & 0.1     & 0.11    & 0.18   & 0.1     & 0.15   \\ \hline
\multicolumn{1}{|l|}{initial\_meanbeta}           & 255150 & 0.02    & 0.00259 & 0.00363 & 0.01    & 0.02    & 0.02    & 0.06   & 0.02    & 0.06   \\ \hline
\multicolumn{1}{|l|}{initial\_meanalphaonbeta}    & 255150 & 11      & 0.82    & 4.66    & 10.9    & 11      & 11.1    & 17.35  & 11      & 12.69  \\ \hline
\multicolumn{1}{|l|}{initial\_bclone1}            & 255150 & 499.96  & 288.48  & 0       & 250     & 500     & 750     & 999    & 500     & 999    \\ \hline
\multicolumn{1}{|l|}{initial\_bclone2}            & 255150 & 499.32  & 288.51  & 0       & 249     & 499     & 749     & 999    & 499     & 999    \\ \hline
\multicolumn{1}{|l|}{initial\_bclone3}            & 255150 & 499.25  & 288.74  & 0       & 249     & 499     & 749     & 999    & 499     & 999    \\ \hline
\multicolumn{1}{|l|}{initial\_bclone4}            & 255150 & 498.44  & 288.18  & 0       & 249     & 498     & 747     & 999    & 498     & 999    \\ \hline
\multicolumn{1}{|l|}{initial\_bclone5}            & 255150 & 500.29  & 288.47  & 0       & 251     & 501     & 749     & 999    & 501     & 999    \\ \hline
\multicolumn{1}{|l|}{initial\_bclone1perc}        & 255150 & 4.92    & 8.1     & 0.12    & 0.22    & 0.73    & 4.37    & 69.88  & 0.73    & 69.76  \\ \hline
\multicolumn{1}{|l|}{initial\_bclone2perc}        & 255150 & 3.33    & 4.99    & 0.12    & 0.21    & 0.64    & 3.35    & 32.72  & 0.64    & 32.59  \\ \hline
\multicolumn{1}{|l|}{initial\_bclone3perc}        & 255150 & 2.61    & 3.69    & 0.12    & 0.2     & 0.59    & 2.88    & 21.11  & 0.59    & 20.99  \\ \hline
\multicolumn{1}{|l|}{initial\_bclone4perc}        & 255150 & 2.17    & 2.92    & 0.12    & 0.2     & 0.56    & 2.57    & 16.26  & 0.56    & 16.14  \\ \hline
\multicolumn{1}{|l|}{initial\_bclone5perc}        & 255150 & 1.86    & 2.4     & 0.12    & 0.2     & 0.54    & 2.35    & 12.82  & 0.54    & 12.7   \\ \hline
\multicolumn{1}{|l|}{initial\_nclone99}           & 255150 & 634.72  & 388.28  & 17      & 298     & 863     & 977     & 988    & 863     & 971    \\ \hline
\multicolumn{1}{|l|}{D\_RC1}                      & 255150 & 72      & 9.8     & 60      & 60      & 72      & 84      & 84     & 72      & 24     \\ \hline
\multicolumn{1}{|l|}{d\_RC1}                      & 255150 & 2.33    & 1.25    & 1       & 1       & 2       & 4       & 4      & 2       & 3      \\ \hline
\multicolumn{1}{|l|}{Nfract\_RC1}                 & 255150 & 42      & 23.37   & 15      & 21      & 36      & 60      & 84     & 36      & 69     \\ \hline
\multicolumn{1}{|l|}{interval\_RC1}               & 255150 & 1.17    & 0.62    & 0.5     & 0.5     & 1       & 2       & 2      & 1       & 1.5    \\ \hline
\multicolumn{1}{|l|}{radio\_meanprol\_RC1}        & 255150 & 0.05    & 0.00166 & 0.03    & 0.05    & 0.05    & 0.05    & 0.05   & 0.05    & 0.02   \\ \hline
\multicolumn{1}{|l|}{radio\_meanalpha\_RC1}       & 255150 & 0.02    & 0.00765 & 0.01    & 0.01    & 0.02    & 0.02    & 0.11   & 0.02    & 0.1    \\ \hline
\multicolumn{1}{|l|}{radio\_meanbeta\_RC1}        & 255150 & 0.00216 & 0.00115 & 0.0005  & 0.00148 & 0.00183 & 0.00246 & 0.03   & 0.00183 & 0.03   \\ \hline
\multicolumn{1}{|l|}{radio\_meanalphaonbeta\_RC1} & 255150 & 11.97   & 2.14    & 2       & 11.38   & 11.97   & 12.69   & 20     & 11.97   & 18     \\ \hline
\multicolumn{1}{|l|}{radio\_bclone1\_RC1}         & 255150 & 343.36  & 258.06  & 6       & 109     & 309     & 509     & 949    & 309     & 943    \\ \hline
\multicolumn{1}{|l|}{radio\_bclone2\_RC1}         & 255150 & 360.58  & 260.99  & 4       & 109     & 309     & 549     & 949    & 309     & 945    \\ \hline
\multicolumn{1}{|l|}{radio\_bclone3\_RC1}         & 255150 & 371.9   & 261.14  & 4       & 118     & 309     & 609     & 949    & 309     & 945    \\ \hline
\multicolumn{1}{|l|}{radio\_bclone4\_RC1}         & 255150 & 382.67  & 261.14  & 4       & 139     & 338     & 609     & 959    & 338     & 955    \\ \hline
\multicolumn{1}{|l|}{radio\_bclone5\_RC1}         & 255150 & 393.56  & 260.29  & 4       & 209     & 409     & 609     & 959    & 409     & 955    \\ \hline
\multicolumn{1}{|l|}{radio\_bclone1perc\_RC1}     & 255150 & 28.23   & 24.64   & 3.86    & 10.08   & 16.83   & 39.54   & 99.99  & 16.83   & 96.13  \\ \hline
\multicolumn{1}{|l|}{radio\_bclone2perc\_RC1}     & 255150 & 12.32   & 7.12    & 0.00066 & 8.13    & 10.09   & 14.85   & 49.68  & 10.09   & 49.68  \\ \hline
\multicolumn{1}{|l|}{radio\_bclone3perc\_RC1}     & 255150 & 8.19    & 3.73    & 0.00056 & 6.42    & 8.23    & 9.81    & 32.38  & 8.23    & 32.38  \\ \hline
\multicolumn{1}{|l|}{radio\_bclone4perc\_RC1}     & 255150 & 6.37    & 2.9     & 0.00055 & 4.71    & 7.04    & 8.26    & 22.1   & 7.04    & 22.1   \\ \hline
\multicolumn{1}{|l|}{radio\_bclone5perc\_RC1}     & 255150 & 5.32    & 2.65    & 0.00042 & 3.41    & 6.01    & 7.4     & 16.11  & 6.01    & 16.11  \\ \hline
\multicolumn{1}{|l|}{radio\_nclone99\_RC1}        & 255150 & 234.76  & 312.24  & 1       & 43      & 66      & 471     & 906    & 66      & 905    \\ \hline
\multicolumn{1}{|l|}{evol\_meanprol\_RC1}         & 255150 & 0.05    & 0.00149 & 0.04    & 0.05    & 0.05    & 0.05    & 0.05   & 0.05    & 0.01   \\ \hline
\multicolumn{1}{|l|}{evol\_meanalpha\_RC1}        & 255150 & 0.1     & 0.02    & 0.01    & 0.1     & 0.1     & 0.11    & 0.2    & 0.1     & 0.19   \\ \hline
\multicolumn{1}{|l|}{evol\_meanbeta\_RC1}         & 255150 & 0.02    & 0.00487 & 0.00097 & 0.01    & 0.02    & 0.02    & 0.1    & 0.02    & 0.1    \\ \hline
\multicolumn{1}{|l|}{evol\_meanalphaonbeta\_RC1}  & 255150 & 11      & 1.52    & 2.17    & 10.77   & 11      & 11.22   & 19.89  & 11      & 17.72  \\ \hline
\multicolumn{1}{|l|}{evol\_bclone1\_RC1}          & 255150 & 503.29  & 288.9   & 7       & 249     & 499     & 749     & 999    & 499     & 992    \\ \hline
\multicolumn{1}{|l|}{evol\_bclone2\_RC1}          & 255150 & 504.06  & 288.26  & 6       & 259     & 508     & 749     & 999    & 508     & 993    \\ \hline
\multicolumn{1}{|l|}{evol\_bclone3\_RC1}          & 255150 & 504.18  & 288.92  & 7       & 249     & 509     & 758     & 999    & 509     & 992    \\ \hline
\multicolumn{1}{|l|}{evol\_bclone4\_RC1}          & 255150 & 503.97  & 288.84  & 6       & 249     & 508     & 757     & 999    & 508     & 993    \\ \hline
\multicolumn{1}{|l|}{evol\_bclone5\_RC1}          & 255150 & 504.35  & 288.63  & 6       & 259     & 507     & 758     & 999    & 507     & 993    \\ \hline
\multicolumn{1}{|l|}{evol\_bclone1perc\_RC1}      & 255150 & 13.31   & 18.78   & 0.7     & 1.41    & 3.83    & 17.45   & 98.8   & 3.83    & 98.1   \\ \hline
\multicolumn{1}{|l|}{evol\_bclone2perc\_RC1}      & 255150 & 6.64    & 7.63    & 0.25    & 1.32    & 3.15    & 9.97    & 48.24  & 3.15    & 48     \\ \hline
\multicolumn{1}{|l|}{evol\_bclone3perc\_RC1}      & 255150 & 4.39    & 4.27    & 0.1     & 1.26    & 2.73    & 6.83    & 30.16  & 2.73    & 30.06  \\ \hline
\multicolumn{1}{|l|}{evol\_bclone4perc\_RC1}      & 255150 & 3.27    & 2.77    & 0.07    & 1.2     & 2.39    & 4.94    & 21.4   & 2.39    & 21.33  \\ \hline
\multicolumn{1}{|l|}{evol\_bclone5perc\_RC1}      & 255150 & 2.62    & 1.97    & 0.04    & 1.01    & 2.03    & 3.62    & 15.55  & 2.03    & 15.51  \\ \hline
\multicolumn{1}{|l|}{evol\_nclone99\_RC1}         & 255150 & 234.76  & 312.24  & 1       & 43      & 66      & 471     & 906    & 66      & 905    \\ \hline
\multicolumn{1}{|l|}{detection\_RC1}              & 255150 & 406.77  & 6.74    & 372.13  & 403.13  & 406.88  & 407.83  & 479.96 & 406.88  & 107.83 \\ \hline
\multicolumn{1}{|l|}{killedperc\_RC1}             & 255150 & 76.01   & 22.75   & -1.22   & 63.27   & 84.98   & 92.45   & 100    & 84.98   & 101.22 \\ \hline
\multicolumn{1}{|l|}{D\_RC2}                      & 255150 & 72      & 9.8     & 60      & 60      & 72      & 84      & 84     & 72      & 24     \\ \hline
\multicolumn{1}{|l|}{d\_RC2}                      & 255150 & 2.33    & 1.25    & 1       & 1       & 2       & 4       & 4      & 2       & 3      \\ \hline
\multicolumn{1}{|l|}{Nfract\_RC2}                 & 255150 & 42      & 23.37   & 15      & 21      & 36      & 60      & 84     & 36      & 69     \\ \hline
\multicolumn{1}{|l|}{interval\_RC2}               & 255150 & 1.17    & 0.62    & 0.5     & 0.5     & 1       & 2       & 2      & 1       & 1.5    \\ \hline
\multicolumn{1}{|l|}{radio\_meanprol\_RC2}        & 255150 & 0.05    & 0.00152 & 0.03    & 0.05    & 0.05    & 0.05    & 0.05   & 0.05    & 0.02   \\ \hline
\end{tabular}
\caption{Descriptive statistics of the dataset (part 1).}
\label{tab:1}
\end{table*}
\begin{table*}[]
\begin{tabular}{lllllllllll}
\hline
\multicolumn{1}{|l|}{evol\_bclone2perc\_RC2} & \multicolumn{1}{l|}{255150}   & \multicolumn{1}{l|}{12.32}     & \multicolumn{1}{l|}{7.12}  & \multicolumn{1}{l|}{0.00066} & \multicolumn{1}{l|}{8.13}    & \multicolumn{1}{l|}{10.09}  & \multicolumn{1}{l|}{14.85}  & \multicolumn{1}{l|}{49.68}  & \multicolumn{1}{l|}{10.09}  & \multicolumn{1}{l|}{49.68}  \\ \hline
\multicolumn{1}{|l|}{evol\_bclone3perc\_RC2} & \multicolumn{1}{l|}{255150}   & \multicolumn{1}{l|}{8.19}      & \multicolumn{1}{l|}{3.73}  & \multicolumn{1}{l|}{0.00056} & \multicolumn{1}{l|}{6.42}    & \multicolumn{1}{l|}{8.23}   & \multicolumn{1}{l|}{9.81}   & \multicolumn{1}{l|}{32.38}  & \multicolumn{1}{l|}{8.23}   & \multicolumn{1}{l|}{32.38}  \\ \hline
\multicolumn{1}{|l|}{evol\_bclone4perc\_RC2} & \multicolumn{1}{l|}{255150}   & \multicolumn{1}{l|}{6.37}      & \multicolumn{1}{l|}{2.9}   & \multicolumn{1}{l|}{0.00055} & \multicolumn{1}{l|}{4.71}    & \multicolumn{1}{l|}{7.04}   & \multicolumn{1}{l|}{8.26}   & \multicolumn{1}{l|}{22.1}   & \multicolumn{1}{l|}{7.04}   & \multicolumn{1}{l|}{22.1}   \\ \hline
\multicolumn{1}{|l|}{evol\_bclone5perc\_RC2} & \multicolumn{1}{l|}{255150}   & \multicolumn{1}{l|}{5.32}      & \multicolumn{1}{l|}{2.65}  & \multicolumn{1}{l|}{0.00042} & \multicolumn{1}{l|}{3.41}    & \multicolumn{1}{l|}{6.01}   & \multicolumn{1}{l|}{7.4}    & \multicolumn{1}{l|}{16.11}  & \multicolumn{1}{l|}{6.01}   & \multicolumn{1}{l|}{16.11}  \\ \hline
\multicolumn{1}{|l|}{evol\_nclone99\_RC2}    & \multicolumn{1}{l|}{255150}   & \multicolumn{1}{l|}{162.72}    & \multicolumn{1}{l|}{269.3} & \multicolumn{1}{l|}{1}       & \multicolumn{1}{l|}{25}      & \multicolumn{1}{l|}{34}     & \multicolumn{1}{l|}{97}     & \multicolumn{1}{l|}{895}    & \multicolumn{1}{l|}{34}     & \multicolumn{1}{l|}{894}    \\ \hline
\multicolumn{1}{|l|}{detection\_RC2}         & \multicolumn{1}{l|}{255150}   & \multicolumn{1}{l|}{575.64}    & \multicolumn{1}{l|}{28.63} & \multicolumn{1}{l|}{467.94}  & \multicolumn{1}{l|}{560.9}   & \multicolumn{1}{l|}{570.43} & \multicolumn{1}{l|}{586.48} & \multicolumn{1}{l|}{783.86} & \multicolumn{1}{l|}{570.43} & \multicolumn{1}{l|}{315.92} \\ \hline
\multicolumn{1}{|l|}{killedperc\_RC2}        & \multicolumn{1}{l|}{255150}   & \multicolumn{1}{l|}{43.17}     & \multicolumn{1}{l|}{18.82} & \multicolumn{1}{l|}{-18.39}  & \multicolumn{1}{l|}{26.49}   & \multicolumn{1}{l|}{44}     & \multicolumn{1}{l|}{56.75}  & \multicolumn{1}{l|}{99.83}  & \multicolumn{1}{l|}{44}     & \multicolumn{1}{l|}{118.22} \\ \hline
\multicolumn{1}{|l|}{last\_detection}        & \multicolumn{1}{l|}{255150}   & \multicolumn{1}{l|}{709.08}    & \multicolumn{1}{l|}{50.33} & \multicolumn{1}{l|}{544.75}  & \multicolumn{1}{l|}{679.05}  & \multicolumn{1}{l|}{699.67} & \multicolumn{1}{l|}{733.86} & \multicolumn{1}{l|}{1094.7} & \multicolumn{1}{l|}{699.67} & \multicolumn{1}{l|}{549.95} \\ \hline
\multicolumn{1}{|l|}{relapse\_time\_RC1}     & \multicolumn{1}{l|}{255150}   & \multicolumn{1}{l|}{168.88}    & \multicolumn{1}{l|}{26.17} & \multicolumn{1}{l|}{71.05}   & \multicolumn{1}{l|}{154.86}  & \multicolumn{1}{l|}{164.09} & \multicolumn{1}{l|}{178.96} & \multicolumn{1}{l|}{355.28} & \multicolumn{1}{l|}{164.09} & \multicolumn{1}{l|}{284.23} \\ \hline
\multicolumn{1}{|l|}{relapse\_time\_RC2}     & \multicolumn{1}{l|}{255150}   & \multicolumn{1}{l|}{133.43}    & \multicolumn{1}{l|}{33.72} & \multicolumn{1}{l|}{48.62}   & \multicolumn{1}{l|}{112.31}  & \multicolumn{1}{l|}{123.85} & \multicolumn{1}{l|}{144.96} & \multicolumn{1}{l|}{369.14} & \multicolumn{1}{l|}{123.85} & \multicolumn{1}{l|}{320.52} \\ \hline
\multicolumn{1}{|l|}{selector\_RC1}          & \multicolumn{1}{l|}{255150}   & \multicolumn{1}{l|}{0.18}      & \multicolumn{1}{l|}{0.09}  & \multicolumn{1}{l|}{0.05}    & \multicolumn{1}{l|}{0.13}    & \multicolumn{1}{l|}{0.15}   & \multicolumn{1}{l|}{0.2}    & \multicolumn{1}{l|}{0.96}   & \multicolumn{1}{l|}{0.15}   & \multicolumn{1}{l|}{0.91}   \\ \hline
\multicolumn{1}{|l|}{toxicity\_RC1}          & \multicolumn{1}{l|}{255150}   & \multicolumn{1}{l|}{81.42}     & \multicolumn{1}{l|}{40.95} & \multicolumn{1}{l|}{23.02}   & \multicolumn{1}{l|}{48.9}    & \multicolumn{1}{l|}{75.7}   & \multicolumn{1}{l|}{115.27} & \multicolumn{1}{l|}{170.3}  & \multicolumn{1}{l|}{75.7}   & \multicolumn{1}{l|}{147.28} \\ \hline
\multicolumn{1}{|l|}{TTP\_RC1}               & \multicolumn{1}{l|}{255150}   & \multicolumn{1}{l|}{119.88}    & \multicolumn{1}{l|}{32.77} & \multicolumn{1}{l|}{-1.96}   & \multicolumn{1}{l|}{103.17}  & \multicolumn{1}{l|}{125.32} & \multicolumn{1}{l|}{146.68} & \multicolumn{1}{l|}{316.28} & \multicolumn{1}{l|}{125.32} & \multicolumn{1}{l|}{318.24} \\ \hline
\multicolumn{1}{|l|}{ind1\_RC1}              & \multicolumn{1}{l|}{255150}   & \multicolumn{1}{l|}{231.79}    & \multicolumn{1}{l|}{62.79} & \multicolumn{1}{l|}{-6.17}   & \multicolumn{1}{l|}{200.92}  & \multicolumn{1}{l|}{239.85} & \multicolumn{1}{l|}{279.84} & \multicolumn{1}{l|}{617.34} & \multicolumn{1}{l|}{239.85} & \multicolumn{1}{l|}{623.51} \\ \hline
\multicolumn{1}{|l|}{ind2\_RC1}              & \multicolumn{1}{l|}{255150}   & \multicolumn{1}{l|}{158.69}    & \multicolumn{1}{l|}{47.46} & \multicolumn{1}{l|}{-27.28}  & \multicolumn{1}{l|}{128.17}  & \multicolumn{1}{l|}{156.49} & \multicolumn{1}{l|}{183.41} & \multicolumn{1}{l|}{523.13} & \multicolumn{1}{l|}{156.49} & \multicolumn{1}{l|}{550.41} \\ \hline
\multicolumn{1}{|l|}{ind3\_RC1}              & \multicolumn{1}{l|}{255150}   & \multicolumn{1}{l|}{-42.61}    & \multicolumn{1}{l|}{63.71} & \multicolumn{1}{l|}{-256.38} & \multicolumn{1}{l|}{-88.8}   & \multicolumn{1}{l|}{-25.64} & \multicolumn{1}{l|}{5.91}   & \multicolumn{1}{l|}{170.33} & \multicolumn{1}{l|}{-25.64} & \multicolumn{1}{l|}{426.71} \\ \hline
\multicolumn{1}{|l|}{selector\_RC2}          & \multicolumn{1}{l|}{255150}   & \multicolumn{1}{l|}{0.72}      & \multicolumn{1}{l|}{0.13}  & \multicolumn{1}{l|}{0.11}    & \multicolumn{1}{l|}{0.65}    & \multicolumn{1}{l|}{0.73}   & \multicolumn{1}{l|}{0.8}    & \multicolumn{1}{l|}{1.45}   & \multicolumn{1}{l|}{0.73}   & \multicolumn{1}{l|}{1.34}   \\ \hline
\multicolumn{1}{|l|}{toxicity\_RC2}          & \multicolumn{1}{l|}{255150}   & \multicolumn{1}{l|}{81.42}     & \multicolumn{1}{l|}{40.95} & \multicolumn{1}{l|}{23.02}   & \multicolumn{1}{l|}{48.9}    & \multicolumn{1}{l|}{75.7}   & \multicolumn{1}{l|}{115.27} & \multicolumn{1}{l|}{170.3}  & \multicolumn{1}{l|}{75.7}   & \multicolumn{1}{l|}{147.28} \\ \hline
\multicolumn{1}{|l|}{TTP\_RC2}               & \multicolumn{1}{l|}{255150}   & \multicolumn{1}{l|}{84.43}     & \multicolumn{1}{l|}{25.57} & \multicolumn{1}{l|}{-1.96}   & \multicolumn{1}{l|}{61.58}   & \multicolumn{1}{l|}{82.12}  & \multicolumn{1}{l|}{102.8}  & \multicolumn{1}{l|}{286.6}  & \multicolumn{1}{l|}{82.12}  & \multicolumn{1}{l|}{288.56} \\ \hline
\multicolumn{1}{|l|}{ind1\_RC2}              & \multicolumn{1}{l|}{255150}   & \multicolumn{1}{l|}{161.45}    & \multicolumn{1}{l|}{48.96} & \multicolumn{1}{l|}{-6.03}   & \multicolumn{1}{l|}{118.61}  & \multicolumn{1}{l|}{157.68} & \multicolumn{1}{l|}{193.26} & \multicolumn{1}{l|}{557.16} & \multicolumn{1}{l|}{157.68} & \multicolumn{1}{l|}{563.19} \\ \hline
\multicolumn{1}{|l|}{ind2\_RC2}              & \multicolumn{1}{l|}{255150}   & \multicolumn{1}{l|}{88.89}     & \multicolumn{1}{l|}{43.88} & \multicolumn{1}{l|}{-50.12}  & \multicolumn{1}{l|}{59.36}   & \multicolumn{1}{l|}{88.01}  & \multicolumn{1}{l|}{110.78} & \multicolumn{1}{l|}{467.22} & \multicolumn{1}{l|}{88.01}  & \multicolumn{1}{l|}{517.34} \\ \hline
\multicolumn{1}{|l|}{ind3\_RC2}              & \multicolumn{1}{l|}{255150}   & \multicolumn{1}{l|}{-76.97}    & \multicolumn{1}{l|}{70.64} & \multicolumn{1}{l|}{-278.18} & \multicolumn{1}{l|}{-130.38} & \multicolumn{1}{l|}{-60.7}  & \multicolumn{1}{l|}{-21.91} & \multicolumn{1}{l|}{177.8}  & \multicolumn{1}{l|}{-60.7}  & \multicolumn{1}{l|}{455.98} \\ \hline
                                             &                               &                                &                            &                              &                              &                             &                             &                             &                             &                             \\ \cline{2-3}
\multicolumn{1}{l|}{}                        & \multicolumn{1}{l|}{category} & \multicolumn{1}{l|}{frequency} &                            &                              &                              &                             &                             &                             &                             &                             \\ \cline{1-3}
\multicolumn{1}{|l|}{timing\_RC1}            & \multicolumn{1}{l|}{S}        & \multicolumn{1}{l|}{85050}     &                            &                              &                              &                             &                             &                             &                             &                             \\ \cline{1-3}
\multicolumn{1}{|l|}{timing\_RC1}            & \multicolumn{1}{l|}{P}        & \multicolumn{1}{l|}{85050}     &                            &                              &                              &                             &                             &                             &                             &                             \\ \cline{1-3}
\multicolumn{1}{|l|}{timing\_RC1}            & \multicolumn{1}{l|}{A}        & \multicolumn{1}{l|}{85050}     &                            &                              &                              &                             &                             &                             &                             &                             \\ \cline{1-3}
\multicolumn{1}{|l|}{fractionation\_RC1}     & \multicolumn{1}{l|}{R}        & \multicolumn{1}{l|}{85050}     &                            &                              &                              &                             &                             &                             &                             &                             \\ \cline{1-3}
\multicolumn{1}{|l|}{fractionation\_RC1}     & \multicolumn{1}{l|}{O}        & \multicolumn{1}{l|}{85050}     &                            &                              &                              &                             &                             &                             &                             &                             \\ \cline{1-3}
\multicolumn{1}{|l|}{fractionation\_RC1}     & \multicolumn{1}{l|}{E}        & \multicolumn{1}{l|}{85050}     &                            &                              &                              &                             &                             &                             &                             &                             \\ \cline{1-3}
\multicolumn{1}{|l|}{timing\_RC2}            & \multicolumn{1}{l|}{S}        & \multicolumn{1}{l|}{85050}     &                            &                              &                              &                             &                             &                             &                             &                             \\ \cline{1-3}
\multicolumn{1}{|l|}{timing\_RC2}            & \multicolumn{1}{l|}{A}        & \multicolumn{1}{l|}{85050}     &                            &                              &                              &                             &                             &                             &                             &                             \\ \cline{1-3}
\multicolumn{1}{|l|}{timing\_RC2}            & \multicolumn{1}{l|}{P}        & \multicolumn{1}{l|}{85050}     &                            &                              &                              &                             &                             &                             &                             &                             \\ \cline{1-3}
\multicolumn{1}{|l|}{fractionation\_RC2}     & \multicolumn{1}{l|}{O}        & \multicolumn{1}{l|}{85050}     &                            &                              &                              &                             &                             &                             &                             &                             \\ \cline{1-3}
\multicolumn{1}{|l|}{fractionation\_RC2}     & \multicolumn{1}{l|}{R}        & \multicolumn{1}{l|}{85050}     &                            &                              &                              &                             &                             &                             &                             &                             \\ \cline{1-3}
\multicolumn{1}{|l|}{fractionation\_RC2}     & \multicolumn{1}{l|}{E}        & \multicolumn{1}{l|}{85050}     &                            &                              &                              &                             &                             &                             &                             &                             \\ \cline{1-3}
\end{tabular}
\caption{Descriptive statistics of the dataset (part 2). "unif\_rate": uniformity parameter of the distribution among all clones of the initial population; 
"mut\_rate":$\mu$.
"Initial","evol","radio" refers to the time of evaluation: initial time, evolutive dynamics without radiotherapy, evolutive dynamics after radiotherapy. "RC\{$i$\} refers to the radiotherapy cycle: 1 first-line, 2 second-line.
"meanprol":$\bar{\gamma}$;
"meanalpha":$\bar{\alpha}$;
"meanbeta:$\bar{\beta}$;
"meanalphaonbeta:$\bar{\frac{\alpha}{\beta}}$.
"bclone\{$i$\}" for $i=1,...,5$: index of the five most abundant clones among the list of indices of the subpopulations; "bclone\{$i$\}perc" for $i=1,...,5$: percentage on the total mass of the five most abundant clones, "nclone99": minimum number of clones representing the 99$\%$ of the mass. "D": total dose, "d" single dose,"Nfract" number of doses, "interval" time in days between the doses.
"detection": progressive time of detection in days;
"killedperc": difference between population size at detection and at the end of the treatment divided by the size at detection.
"last\_detection": progressive time of last detection.  
"selector": ratio between $\bar{alpha}$ after and before radiotherapy,
"toxicity": toxicity indicator estimated with the damage to healthy cells considering repair between the doses;
"relapse\_time": time in day between two detections, 
"TTP": time to progression, equal to the difference between the relapse time and the duration of the treatment (can be negative if the tumour mass reaches the detection size during the treatment because off radiotherapy failure);
"ind\{$i$\}" for $i=1,...,3$ indicators of treatment performances obtained by weighted linear combination of "selector", "toxicity" and "TTP". "timing": indicator of the chosen timing of the protocol: S-standard (one day interval), P-protracted (two days interval), A-accelerated (half day interval); "fractionation": indicator of the chosen fractionation of the protocol: R-regular (2 Gy dose), E-hyperfractionated (1 Gy dose), O-hypofractionated (4 Gy dose).}
\label{tab:2}
\end{table*}

\end{document}